\documentclass{amsart}
\usepackage{graphicx}  
\usepackage{multicol}  
\usepackage{amsmath}   
\usepackage{amsfonts}   
\usepackage{amssymb}
\usepackage{url}

\DeclareFontFamily{OT1}{rsfs}{}
\DeclareFontShape{OT1}{rsfs}{m}{n}{ <-7> rsfs5 <7-10> rsfs7 <10->rsfs10}{}
\DeclareMathAlphabet{\mycal}{OT1}{rsfs}{m}{n}
\def\scri{{\mycal I}}%

\newcounter{mnotecount}[section]%

\let\index\relax

\begin{document}

\newcommand{\aM}{\mathbf M}
\newcommand{\ame}{\mathbf g}
\newcommand{\aR}{\mathbf R}
\newcommand{\anabla}{\mathbf D}
\newcommand{\aC}{\mathbf C}
\newcommand{\taM}{{\mathbf {\tilde \aM}}}
\newcommand{\tame}{\mathbf {\tilde \ame}}
\newcommand{\taR}{{\mathbf {\tilde \aR}}}
\newcommand{\tanabla}{{\mathbf {\tilde \anabla}}}
\newcommand{\bOmega}{\bar \Omega}
\newcommand{\bame}{\mathbf {\bar \ame}}
\newcommand{\baM}{{\mathbf {\bar \aM}}}
\newcommand{\bM}{\bar M}
\newcommand{\norm}{\eta}        
\newcommand{\Lapse}{N}
\newcommand{\Shift}{X}
\newcommand{\la}{\langle}               
\newcommand{\ra}{\rangle}               
\newcommand{\half}{\frac{1}{2}}         
\newcommand{\third}{\frac{1}{3}}        
\newcommand{\Lie}{\mathcal L}
\newcommand{\PM}{\big{|}_{\partial M}}
\newcommand{\inabla}{\nabla \! \! \!\!  / \,\,}
\newcommand{\iR}{R \!\!\!\! / \,\,}
\newcommand{\tK}{\tilde K}
\def\tg{\tilde g}
\def\Re{{\mathbb{R}}}
\newcommand{\tLapse}{\tilde \Lapse}
\newcommand{\tShift}{\tilde \Shift}
\newcommand{\tnabla}{\tilde \nabla}
\newcommand{\tR}{\tilde R}
\newcommand{\tT}{\tilde T}
\newcommand{\taC}{{\mathbf {\tilde \aC}}}
\newcommand{\tr}{\text{\rm tr}}         
\newcommand{\ttr}{\widetilde{\tr}}
\newcommand{\hme}{\hat g}
\newcommand{\hnabla}{\hat \nabla}
\newcommand{\hR}{\hat R}
\newcommand{\hL}{\hat L}
\newcommand{\hDelta}{\hat \Delta}
\newcommand{\hsigma}{\hat \sigma}
\newcommand{\cC}{\mathcal C}
\newcommand{\eps}{\epsilon}

\title{Construction of hyperboloidal initial data}
\author{Lars Andersson}
\thanks{This paper is based on a talk given at the Workshop on the Conformal
Structure of Space-Times held at T\"ubingen, April 2--4, 2001. I am
grateful to the organizers, J\"org Frauendiener and Helmut Friedrich,
for their hospitality and support. This work is supported in part by
the Swedish Natural Sciences Research Council (SNSRC), contract no.
R-RA 4873-307 and NSF, contract no. DMS 0104402.}
\address{Department of Mathematics, University of Miami, Coral
  Gables, FL 33124, USA}

\allowdisplaybreaks[2]

\maketitle


\section{Introduction}
Let $(\aM, \ame, \Omega)$ be the conformal rescaling of a 3+1 dimensional 
vacuum
spacetime $(\taM, \tame)$, with $\ame = \Omega^2 \tame$, where
$\Omega$ is a smooth function on $\aM$ vanishing on the boundary
$\partial \aM$. If $\aM$ has smooth null boundary $\scri$, 
with connected components of topology $S^2 \times \Re$,
it follows that the Weyl tensor of $(\aM, \ame)$ vanishes on
$\scri$. From this follows peeling properties for $(\taM,
\tame)$. This picture of isolated systems in general relativity,
developed by Penrose,
see~\cite{penrose:rindler:II,frauendiener:living}, is useful for
studying the mass and angular momentum of spacetimes, as well as
gravitational radiation.

Friedrich has found a first order symmetric--hyperbolic version of the 
Einstein evolution equations, called the conformally regular field equations,
which may be extended through $\scri$ \cite{HF}. 
The Cauchy data for this system on a
future Cauchy surface $M$, asymptotic to $\scri$ so that 
$\bar M \cap \partial \aM = \partial M$, 
include $g_{ab}, K_{ab},
\Omega$ as well as components of the rescaled Weyl tensor
$\Omega^{-1}\aC^\alpha_{\ \beta\gamma\delta}$. In order to get a
regular evolution at $\scri$, these data must be regular up to
$\partial M$. It was shown by the author, Chrusciel and
Friedrich~\cite{ACF,AC,ACPRL,ACdiss} that under certain conditions on
the boundary geometry of $M$, the Cauchy data for the conformally
regular field equations, are smooth up to $\partial M$.

Using the conformally regular field equations, Friedrich has shown
that the maximal vacuum development of data (called hyperboloidal data) 
on a future Cauchy surface intersecting $\scri$, regular up to $\partial M$ 
is a
spacetime which has a ``smooth piece of $\scri$''. Further, for small
data the maximal vacuum development has a null boundary with future
complete null generators and a regular timelike infinity.

A programme has been initiated to numerically evolve the Einstein equations
using the conformally regular field equations~\cite{hubner1,hubner2,hubner3,hubner4}. 
This approach may have advantages
over the ``traditional'' approach 
which treats the asymptotic flatness condition
by introducing boundary conditions far away from the isolated system under
study, and attempts to observe for example gravitational wave signatures on
this boundary. 

In this note we will discuss the conformal procedure for constructing
solutions $(g_{ab}, K_{ab}, \Omega)$ 
to the constraint equations and 
the geometric conditions for regularity
at $\scri$. I will also briefly discuss the Cauchy problem for the Einstein
evolution equations at $\scri$, directly from the point of view of the
Einstein equations in $(\aM, \ame)$.

\section{Preliminaries}\label{sec:prel}
The following index conventions will be used. Greek indices
$\alpha,\beta,\dots$ run
over $0,\dots,3$, lower case latin indices 
$a,b,c,\dots$ run over $1,2,3$, and  upper case latin indices $A,B,C,\dots$ run
over $2,3$. 

Let $(\aM,\ame)$ be a 3+1 dimensional 
globally hyperbolic space--time with covariant
derivative $\anabla$. Introduce coordinates $(x^{\alpha}) = (t,x^a)$ on $\aM$, and consider the
foliation $M_t$ consisting of level sets of $t$. We will often drop the
subscript $t$ on $M_t$ and associated fields. $M$ will be a compact manifold
with boundary $\partial M$, and closure $\bM = M \cup \partial M$. 
Let $T$ be the normal of $M$, assumed to be timelike, and let $g$ be the
induced positive definite metric on $M$, with
covariant derivative $\nabla$. We will use an index $T$ to denote contraction
with $T$, for example $t_T = T^{\alpha} t_{\alpha}$. 
Let $\Sigma = \anabla_T \Omega$ and assume $|\Sigma| > 0$. The  
lapse function $\Lapse$ and shift
vectorfield $\Shift$ are defined by $\partial_t = \Lapse T + \Shift$. 
The
second fundamental form $K$ is defined by 
$K_{ab} = \la \anabla_a e_b  , T \ra$. Then we have 
$K_{ab} = - \half \Lie_T \ame_{ab} = 
- \half \Lapse^{-1} (\partial_t - \Shift) g_{ab}$ and $\anabla_a e_b =
\nabla_a e_b - T K_{ab}$.

Assume that $\aM$ has null boundary $\partial \aM$ and
let $\baM = \aM \cup \partial \aM$. Let
$\Omega$ be a positive function on $\aM$ with $\Omega = 0$, $d\Omega \ne 0$ 
on $\partial \aM$ and denote by $\tame$ the conformally
related metric $\tame = \Omega^{-2} \ame$. We will
call $\ame$ the unphysical metric and $\tame$ the
physical metric. We will use $\la \cdot , \cdot \ra$ for the inner product
induced on $T\aM$ by $\ame$. 
Geometric quantities associated with $\tame$
will be decorated with a tilde $\sim$, for example the covariant derivative 
$\tanabla$ and the Ricci tensor $\taR_{\alpha\beta}$.
We will consider only the case when $(\taM, \tame)$ satisfies the
vacuum Einstein equations $\taR_{\alpha\beta} = 0$.
Let $\omega = \Omega \big{|}_{M}$. 
Assume that $\partial M = \bM \cap \partial
\aM$, has positive definite induced metric 
and that $\omega$ is a defining function for $\partial M$, 
i.e. $d\omega \PM \ne 0$. We will refer to $\partial \aM$ as $\scri$ and 
the surface $\partial M \subset
\partial \aM$ as a cross section of $\scri$. 

There is a gauge ambiguity in the conformal rescaling. Let $\bOmega = \Omega
\Theta^{-1}$, $\bame = \Theta^{-2} \ame$ for some positive 
function $\Theta$ which is
bounded  and bounded away from zero on $\aM$. 
Then $\tame = \bOmega^{-2} \bame$ is another conformal rescaling of $\ame$.
A conformal gauge change can be used to control the mean
curvature of $\partial M$, as well as the mean curvature of $M$. 
In particular, the conformal gauge freedom can be used so that the level sets
of $\omega$ are the leaves of the gauss foliation w.r.t. $\partial M$,
i.e. there are coordinates $(x^a) = (x,y^A)$, so that 
in a neighborhood of $\partial M$, we have 
$\omega = x$, and the metric takes the form 
\begin{equation}\label{eq:gform}
g_{ab} dx^a dx^b = dx^2 + h_{AB}dy^A dy^B .
\end{equation}
This possible without loss of generality, cf.~\cite[Lemma 2.1]{ACF}. By
construction 
we have $x(p) = d(p,\partial M)$ and the unit normal 
$\partial M$ is $\norm = \partial_x$. We use the index 1 for contraction with
$\eta$, for example $t_{1b} = \nabla^a x
t_{ab} = \norm^a t_{ab}$. Let $\inabla_A$ and $\iR_{AB}$ 
denote the covariant derivative with
respect to $h_{AB}$ and the Ricci tensor of $h_{AB}$. 
The second fundamental form of $\partial M$ with respect to $\norm$ is
defined by 
$\lambda_{AB} = \la \nabla_A e_B , \norm \ra$. Then we have 
$\nabla_A e_B = \inabla_A e_B + \norm \lambda_{AB}$ and 
$\lambda_{AB} = - \half \Lie_{\norm} h_{AB}$.

The quantities associated to $(g_{ij},K_{ij},\Lapse,T)$ in the
physical space--time are 
\begin{subequations}\label{eq:confdata}
\begin{align}
\tg_{ij} &= \omega^{-2} g_{ij}, & \tK_{ij} &= \omega^{-1} K_{ij} + \Sigma
\tg_{ij}, \\
\tLapse &= \omega^{-1} \Lapse , &
  \tT &= \omega T . 
\end{align}
\end{subequations}
The shift vectorfield $\Shift$ does not scale, so 
$\tShift = \Shift$. The physical 
Weyl tensor satisfies 
$\taC_{\alpha \beta \gamma \delta} = \taR_{\alpha\beta\gamma\delta}$ since 
$\taR_{\alpha\beta} = 0$. Further, 
$\taC^{\alpha}_{\ \beta\gamma\delta} = \aC^{\alpha}_{\ \beta\gamma\delta}$.

A symmetric tensor field $A_{ab}$ on $M$ is said to satisfy the 
\index{shear free condition} if 
\begin{equation}\label{eq:shearfree}
 ( A_{AB} - \half h^{CD} A_{CD} h_{AB}  ) \PM
= \frac{\Sigma}{|\Sigma|} 
 ( \lambda_{AB} - \half h^{CD} \lambda_{CD} h_{AB}  ) 
\PM .
\end{equation}
In the following we will consider only
hypersurfaces intersecting $\scri_+$ which in view of our conventions force
$\Sigma < 0$, so that $\Sigma / |\Sigma| = -1$. 
The spacetime $(\aM, \ame)$ is said to be shear free at $\partial M$ if 
$K_{ab}$ satisfies the shear free condition. 
In order for $\ame$ to be in $C^2(\baM)$, it is necessary that $K_{ab}$
satisfy the shear free condition, cf.~\cite[Prop. 3.1]{AC}. 
In the following we 
only consider initial data which are shear free. 
Note that a change of conformal gauge can be used to get $h^{AB}
\lambda_{AB} \PM = 0$.

\section{Conformal rescalings of Minkowski space}\label{sec:LAconf}
In order to understand the Einstein equations near $\scri$ it is important to
choose an appropriate conformal compactification and a suitable foliation
near $\scri$. In this section we will display a few examples in the simplest
case, Minkowski space. 

Minkowski space is $\Re^{4}$ with the
flat line element 
$$
d\tilde s^2 = - dt^2 + dr^2 + r^2 d\sigma^2 ,
$$
in radial coordinates $(t,r,\theta,\phi)$,
where $d\sigma^2 = d\theta^2 + \sin^2(\theta)d\phi^2$ 
is the line element on $S^2$. 
We will consider two conformal rescalings of Minkowski space. 
First, define new coordinates $\tau, \rho$ by 
$$
t+r = \tan\left(\frac{\tau+\rho}{2}\right), \qquad t-r = 
\tan\left(\frac{\tau-\rho}{2}\right) ,
$$
on
\begin{equation}\label{eq:domain}
U = \{\tau, \rho : 
-\pi < \tau < \pi , \quad 0 \leq \rho < \pi, \quad |\tau+\rho| < \pi, 
\quad 
|\tau-\rho| < \pi \} .
\end{equation}
Rescaling with the conformal factor 
$$
\Omega(\tau,\rho) = 2\cos\left(\frac{\tau+\rho}{2}\right)
\cos\left(\frac{\tau-\rho}{2}\right) ,
$$
which is positive on $U$, 
gives~\cite[p. 450-451]{HF}
$$
ds^2 = \ame_{\alpha\beta} dx^\alpha dx^\beta 
 = -d\tau^2 + d\rho^2 + \sin^2(\rho)d\sigma^2  , 
$$
which is the line element of the Einstein universe 
$\Re \times S^3$. We have $\Lapse \equiv 1$, $\Shift \equiv 0$. 

The null part of the boundary of $U$ has two components with $\tau > 0$ or
$\tau < 0$, called future and
past null infinity, denoted $\scri_{\pm}$.
The point $(\tau,\rho) = (0,\pi)$
is called spatial infinity and denoted $i_0$. The points $i_{\pm}$ with
$(\tau,\rho) = (\pm \pi , 0)$ are called future and past timelike infinity. 
All inextendible future directed null curves start at $\scri_-$ and end at
$\scri_+$.  

\begin{figure}
\centering
\includegraphics[height=2in]{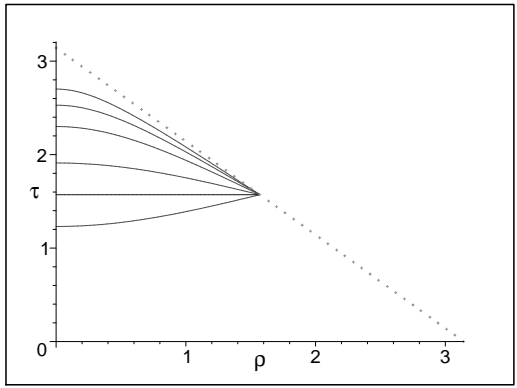}
\caption{Hyperboloids $t^2 - r^2 = u^2$ in conformally compactified Minkowski
space.}
\label{fig:conf}
\end{figure}

\begin{figure}
\centering
\includegraphics[height=2in]{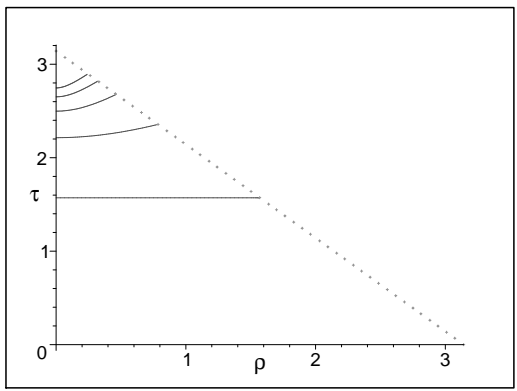}
\caption{Time translated hyperboloids $t = \sqrt{1+r^2} + u$
in conformally compactified Minkowski
space.} 
\label{fig:translatedhyp}
\end{figure}

The hyperboloids $t^2 - r^2 = u^2$ in Minkowski space correspond to 
hypersurfaces in $U$ intersecting $\partial U$ at $\tau = \pi/2$ at different
angles depending on $u$, cf. figure \ref{fig:conf}. In particular, the
hyperboloid with sectional  curvature $-1$ corresponds to $\tau = \pi/2$.
On the other hand, the time translated hyperboloids given by $t =
\sqrt{1+r^2} + u$ intersect $\scri$ at $\tau = \pi/2 + \arctan(u)$, see
figure \ref{fig:translatedhyp}. In this coordinate system, the future 
translated
hyperboloids approach timelike infinity $i_+$ as $u$ increases. The
Lapse $N$ is given at $\scri$ by  
$$
\lim_{r\to \infty} N = \frac{1}{\sqrt{1+u^2}} = \cos(\tau-\pi/2)
$$
which tends to zero as $\tau$ increases to $\pi$. 

For the second conformal rescaling, introduce new coordinates $\psi,\tau$ by 
\begin{align*}
t &=  \coth(\psi) + \tau , \\
r &= 1/\sinh(\psi) .
\end{align*}
This is a singular coordinate transformation, but this does not cause a
problem as we are only interested in the behavior near infinity. 
In terms of the new coordinates, the Minkowski line element takes the form 
$$
d\tilde s^2 = - d\tau^2 + 2\sinh^{-2}(\psi) d\psi d\tau 
+ \sinh^{-2}(\psi) ( d\psi^2 + d\sigma^2) .
$$
After a conformal rescaling with 
$$
\Omega = \sinh(\psi) ,
$$
we get the line element $ds^2 = \Omega^2 d\tilde s^2$ given by 
\begin{equation}\label{eq:confsinhmet}
ds^2 = - \sinh^2(\psi) d\tau^2 +  2  d\psi d\tau 
+ ( d\psi^2 +  d\sigma^2) .
\end{equation}
The hypersurfaces $\tau = $constant correspond to the time translated 
hyperboloids 
$t = \sqrt{1+r^2} + \tau$ in Minkowski space. 
The conformally rescaled spacetime $\aM$ 
has null boundary $\scri_+ = \{ \psi = 0\}$, and so the boundary of this  
conformal rescaling does not include $\scri_-$ or $i_0, i_{\pm}$.
Further, the closure of $\aM$ is not compact.

\begin{figure}
\centering
\includegraphics[height=2in]{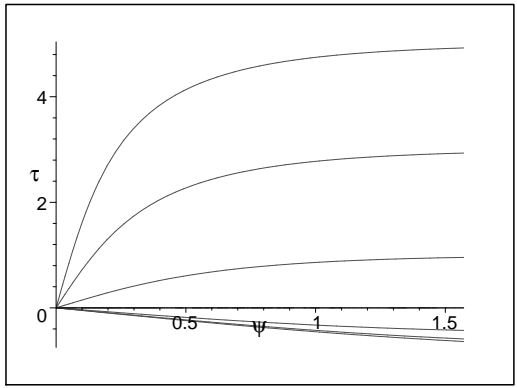}
\caption{Hyperboloids $t^2 - r^2 = u^2$ in the conformal rescaling (\ref{eq:confsinhmet}).}
\label{fig:conf-sinh}
\end{figure}

The foliation $M_\tau$ of level sets of $\tau$ has Lapse 
$\Lapse = \cosh(\psi)$, and Shift $\Shift = \partial_{\psi}$ so that  
the time like normal $T$ is 
$T = \cosh^{-1}(\psi) ( \partial_\tau - \partial_\psi)$. 
The foliation $M_\tau$ is static so that $K_{ab} = 0$. The induced geometry
on $M_\tau$ is the cylinder $d\psi^2 + d\sigma^2$,
with totally geodesic boundary $\{ \psi = 0 \}$. 
The future sheets of the hyperboloids
$t^2 - r^2 = u^2$ of mean curvature $-3/u$ 
intersect $\scri$ at $\tau = 0$, see figure
\ref{fig:conf-sinh}.

\section{Conformal constraint equations} \label{sec:confconstr}
The Einstein vacuum equations $\taR_{\alpha\beta} = 0$ imply the constraint
equations 
\begin{subequations}\label{eq:constraints}
\begin{align}
\tR + (\tK^a_{\ a})^2 - \tK_{ab}\tK^{ab} &= 0 , \\
\tnabla^a \tK_{ab} - \tnabla_b \tK^c_{\ c} &= 0 .
\end{align}
\end{subequations}
We will consider the constraint equations, in the special case of constant
mean curvature $\nabla_a \ttr \tK = 0$, 
as a system of equations for the unphysical data 
and rewrite them 
in a form which exhibits the freely specifiable data. 

In the form of the constraint equations that we will present, 
the freely specifiable data are $(g_{ij}, A_{ij})$, where $g_{ij}$ is a
metric on $M$, and $A_{ij}$ is a trace free, symmetric tensor field on $M$
satisfying the shear free condition. 
We will assume for simplicity 
$A_{1a}\PM = 0$, which does not restrict the degrees of
freedom.  
 
Let $\rho$ be a
defining function for $\partial M$, $\rho =  x$ near $\partial M$. 
Assume $\Sigma \big{|}_{M}$ is a nonzero constant. Set 
$\hme_{ab} = \rho^{-2} g_{ab}$, 
define $\hnabla, \hR$ etc. w.r.t. $\hme$ and raise and lower indices
on hatted fields with $\hme$. 
Given data $g_{ij}, A_{ij},$ a solution $(g_{ij}, K_{ij} , \omega)$ to the
constraint equations is constructed by solving the system
(\ref{eq:confconstr}) below. We will refer to the system
(\ref{eq:confconstr}) as the \index{conformal constraint equations}. 
Existence, 
uniqueness and regularity of solutions to the conformal constraint equations 
has been analyzed in the papers~\cite{ACF,AC}.
For a vector field $Y$, let 
\begin{subequations}\label{eq:confconstr}
\begin{equation}
L(Y)_{ij} = \nabla_i Y_j + \nabla_j Y_i - \frac{2}{3} \nabla_k Y^k g_{ij} .
\end{equation}
For a solution $Y$ to the system
\begin{equation}
\hnabla^i \hL(Y)_{ij} = \hnabla^i (\rho^{-1} A_{ij} ),
\end{equation}
define $\sigma_{ij}$ by 
\begin{equation}
\sigma_{ij} = A_{ij} - \rho^{-1} L(Y)_{ij} .
\end{equation}
Then  $\hsigma_{ij} = \rho^{-1} \sigma_{ij}$ satisfies $\hnabla^a
\hsigma_{ab} = 0$. Let $u$ be the solution to
the semilinear elliptic equation 
\begin{equation}\label{eq:lichner}
- 8 \hDelta u + \hR u - \hsigma_{ij} \hsigma^{ij} u^{-7} + 6 \Sigma^2 u^5 
=  0 .
\end{equation}
\end{subequations}
Finally, define $(g_{ij}, K_{ij} , \omega)$ by 
$\omega = u^{-2} \rho$, $K_{ij} = u^{-4} \sigma_{ij} + \Sigma \tg_{ij} $. 
The fields $(\tg_{ij}, \tK_{ij})$ corresponding to a
solution $g_{ij} , K_{ij}, \omega$ of the conformal constraint equations
(\ref{eq:confconstr}) solve 
the physical constraint equations (\ref{eq:constraints}) with
mean curvature $\tg^{ij} \tK_{ij} = 3 \Sigma$, 
under the constant mean curvature condition $\nabla_a \Sigma = 0$. 

\subsection{Constant mean curvature hypersurfaces}
The existence of complete 
constant mean curvature hypersurfaces in asymptotically flat spacetimes was
studied in~\cite{andersson:iriondo:CMC}. There it was proved that in an
asymptotically Schwarzschild 
spacetime satisfying a natural 
causal condition (domain of dependece of compact sets
is compact), given a cross section of $\scri_+$, and a number $\Sigma < 0$, 
there is a unique \index{CMC hypersurface} intersecting $\scri_+$ at the given cross
section. The proof uses a barrier construction, together with the causality
condition, to prove that the solutions to a sequence of boundary value
problems converges to a complete CMC hypersurface. 

This result indicates that a wide class of 
spacetimes with regular conformal compactification are foliated by CMC
hypersurfaces. The regularity of the height functions of these CMC
hypersurfaces at $\scri$ has not been studied in detail, nor has the
existence proof been carried out in the conformally compactified
setting. This is a natural problem which should be studied further. 

The conditions for regularity of CMC
solutions of the conformal constraint
equations constructed using the procedure discussed above is well
understood, see the discussion in section \ref{sec:regularity} below. 
However, it is not at all well understood how these results
relate to the regularity of CMC foliations in the physical or unphysical
spacetime, evolved from these data, 
nor what spatial gauge conditions are suitable in this situation, see 
section \ref{sec:ivp} below.

\subsection{Degenerate elliptic equations}
The system (\ref{eq:confconstr}) of 
conformal constraint equations is a \index{degenerate
elliptic system} for $(u, Y^i)$. We will briefly describe the regularity
properties of solutions to systems of this type. As an example consider the
Lichnerowicz equation (\ref{eq:lichner}). The principal part of the right
hand side of 
(\ref{eq:lichner}) is essentially of the form 
$$
L = L_x + B ,
$$
where $L_x$ is the ordinary differential operator 
$L_x = x^2 \partial_x^2 + a x\partial_x  + b$, and $B$ is of lower order in
$\partial_x$. In case $(1-a)^2 > 4b$, 
the equation for $L_x x^{\alpha} = 0$ has 
real distinct characteristic roots $\alpha_{\pm}$ and the 
equation $L_x u = f$ has a  solution $u = o(x^{\alpha_-})$ for sufficiently
regular $f = o(x^{\alpha_-})$. 

The operator appearing in the Lichnerowicz equation has critical exponents
$\alpha_- = -1$, $\alpha_+ = 3$. Let $L$ be this operator and consider the
equation $Lu = f$, for $f \in C^{\infty}(\overline M)$. Suppose $f$ is of
the form $f = f_1 x + f_2 x^2 + f_3 x^3$. Then a solution $u = o(x^{-1})$ 
is of the form 
$$
u = 1 + u_1 x + u_2 x^2 + u_{3,1} x^3 \ln x + u_3 x^3 + \text{ higher order
terms} ,
$$ 
where $u_{3,1} = c f_3$ for some explicit constant $c$. This example reflects
the fact, cf.~\cite{ACF,AC,ACdiss} that 
solutions to degenerate elliptic systems are in general nonsmooth at
$\partial M$, instead the general form of the solution has a polylogarithmic
expansion of the form 
$u = \sum u_{i,j} x^i \ln^j x$.
In case the RHS and the coefficients are smooth, the logarithm terms in
$u$ appear first at the critical exponent $\alpha_+$. 

The terms in the expansion of $u$ up to and including the first
logarithm term are computable in terms of the data, which implies 
that explicit geometric conditions
for the regularity of solutions of the conformal constraint equations
(\ref{eq:confconstr}), and hence of the physical spacetime constructed from
these data, can be found. 

\subsection{Regularity of solutions to the conformal constraint equations}
\label{sec:regularity} 
We will now discuss the detailed conditions for
regularity of the conformal data $(g_{ab}, K_{ab}, \Omega)$ at $\partial M$. 
We will assume that free 
data $(g_{ij}, A_{ij}) \in C^{\infty} ( \bM)$ are given and
consider the conditions needed for the solution $(g_{ij}, K_{ij} , \omega)$
of the conformal constraint equations, produced by solving the system 
(\ref{eq:confconstr}), to be in $C^{\infty}(\bM)$. 
 
In the case when $M$
is a moment of time symmetry in the 
unphysical space--time, i.e. $A_{ij} = K_{ij} = 0$, 
there is a simple criterion for
regularity of the solution to the constraint equations. 
Let 
$$
\lambda^*_{AB} = 
 \lambda_{AB} - \half h^{CD} \lambda_{CD} \lambda_{AB} 
$$
denote the trace free part of $\lambda_{AB}$. 
Then
the solution to the conformal constraint equation is in 
$C^{\infty}(\bM)$ if and only if the conformal density $\cC$ 
on $\partial M$, defined by 
$$
\cC = \inabla^A \inabla^B \lambda_{AB} + \lambda^{*AB} R_{AB} - \half h^{CD}
\lambda_{CD} \lambda^{AB} \lambda_{AB}
$$
vanishes. In particular, this holds if $\partial M$ is totally umbilic, i.e. 
if 
$$
( \lambda_{AB} - \half h^{CD} \lambda_{CD} \lambda_{AB} )
\PM = 0 .
$$
In general if this condition does not hold, then the solution has an
expansion in powers of $x$ and $\ln x$ near $\partial M$ and is thus of
finite regularity at $\partial M$. 

In the case where $K_{ij} \ne 0$ and consequently $A_{ij} \ne 0$, 
we make the following simplifying assumptions. 
We consider shear free data $(g_{ij}, A_{ij}) $, such that 
\begin{subequations}\label{eq:bdrygauge} 
\begin{align}
h^{AB} \lambda_{AB} \PM &=0 ,\label{eq:trlambda} \\
A_{1j} \PM &= 0 . \label{eq:A1j}
\end{align}
\end{subequations}
Equation (\ref{eq:trlambda}) 
should be viewed as a conformal gauge condition, 
which can always be satisfied if the data $(g_{ij}, A_{ij})$
is smooth up to boundary. Similarly, equation (\ref{eq:A1j}) may be viewed as a
gauge type condition as it is possible to achieve this by subtracting a term 
$\rho^{-1} L_{ij}$ from $A_{ij}$. 
If (\ref{eq:bdrygauge}) holds, then the fields $\omega, Y^i$ satisfy 
\begin{subequations}\label{eq:asymptform}
\begin{align}
\omega &= |\Sigma| x  + O(x^3) , \\
Y^i &=  O(x^3) ,
\end{align}
\end{subequations}
and $K_{ab}$ is determined up to first order by $A_{ab}$. 
If (\ref{eq:bdrygauge}) holds, 
then the fields 
$\omega, K_{ij} , \omega^{-1} \aC_{\alpha\beta\gamma\delta}$ are in 
$C^{\infty}(\bM)$ if and only if the equations
\begin{subequations}\label{eq:regcond}
\begin{align}
\partial_x K_{AB} - \half h^{CD} \partial_x K_{CD} h_{AB} &= 0  ,
\label{eq:trdotA} \\
\inabla^A \inabla^B \lambda_{AB} + R_{AB} \lambda^{AB} 
&=  0 ,
\label{eq:divdivlam} \\
\inabla^B (\partial_x K_{AB} + \frac{3}{2} \lambda_{CD}\lambda^{CD} h_{AB} )
&= 0    , 
\label{eq:divp}
\end{align}
\end{subequations}
cf.~\cite[Eq. (4.21)--(4.23), (4.33)]{AC},  
hold on $\partial M$. Note that our sign conventions for
$\eps$ and $K_{ab}$ are opposite those of~\cite{AC}. 
The particular form we present
here holds assuming (\ref{eq:bdrygauge}). 
Equation (\ref{eq:trdotA}), which
states that $\partial_x K_{AB}$ is proportional to $h_{AB}$ on $\partial M$, is a
consequence of the vanishing of the Weyl tensor on $\partial M$, while
equations (\ref{eq:divdivlam}) and (\ref{eq:divp}) follow from smoothness of
$K_{ab}$ up to $\partial M$. 
In particular, it can be shown that if $\partial M \cong S^2$, then
\begin{equation}\label{eq:pzero}
\left ( \partial_x K_{AB} + \frac{3}{2} \lambda^{CD}\lambda_{CD} h_{AB}
\right ) \PM = 0 .
\end{equation}

\section{The initial value problem}\label{sec:ivp} 

In this section we will consider the
Einstein evolution equations directly from the point of view of the
unphysical Cauchy data. Since we know from the work of Friedrich~\cite{HF} 
that the
maximal vacuum extension of conformally regular data on $M$ 
has regular conformal boundary near $\partial M$, 
it follows that solutions to a well--posed formulation of the
Einstein evolution equations in the unphysical space--time will have this 
property. Working directly in terms of the unphysical Cauchy data allows one
to relate this problem to the extensive literature on the numerical solution
of the Einstein evolution equations. In particular, it is interesting to
consider gauge choices and hyperbolic reformulations. 

Consider a foliation $M_t$ of $(\aM,\ame)$ with $\partial_t = NT + X$. 
The structure equations of the foliation
imply the 
Einstein evolution equations 
\begin{subequations}\label{eq:evolution}
\begin{align}
\partial_t g_{ij} &= - 2 N K_{ij} + \Lie_X g_{ij} ,\\
\partial_t K_{ij} &= - \nabla_i \nabla_j N + N ( R_{ij} + 
\tr K K_{ij} - 2 K_{im} K^m_{\ j} - \aR_{ij} ) + \Lie_X K_{ij} , \label{eq:dtK}
\end{align}
\end{subequations}
and the constraint equations
\begin{subequations}\label{eq:unph-constr}
\begin{align} 
R - |K|^2 + (\tr K)^2 &= 2 \aR_{TT} + \aR , \\
\nabla_i \tr K - \nabla^j K_{ij}  &= \aR_{Ti} . 
\end{align}
The unphysical Ricci tensor $\aR_{\alpha\beta}$ is given by 
$$
\aR_{\alpha\beta} =  
\taR_{\alpha\beta}
- \Omega^{-1} [ 2 \anabla_{\alpha}\anabla_{\beta} \Omega 
+ \anabla_{\gamma} \anabla^{\gamma} \Omega \ame_{\alpha\beta}] 
+ 3 \Omega^{-2} \anabla_{\gamma} \Omega \anabla^{\gamma} \Omega
\ame_{\alpha\beta} , 
$$
the unphysical Ricci scalar is 
$\aR = \ame^{\alpha\beta} \aR_{\alpha\beta}$, and $|K|^2 = K_{ij} K^{ij}$. 
The condition $\anabla_T \Omega = \Sigma$ may be
viewed as an evolution equation for $\Omega$,
\begin{equation}\label{eq:dtOmega}
\partial_t \Omega = \Lapse \Sigma + X \Omega .
\end{equation}
\end{subequations}

\subsection{Gauge condition at $\partial M$}
It is natural to require $\partial_t \Omega \PM = 0$.  
From (\ref{eq:dtOmega}) and $\omega = |\Sigma|x + h.o.$, 
$\partial_t \Omega \PM = N \Sigma + |\Sigma| \PM$, so 
$\partial_t \Omega \PM = 0$ implies 
$$
\Lapse \PM = \la \Shift , \norm \ra \PM .
$$
In view of this, a natural
boundary condition for $\Lapse$ and $\Shift$ is 
\begin{equation}\label{eq:bdryNX}
\Lapse \PM = 1, \qquad \Shift \PM = \norm, 
\end{equation}
which corresponds to the asymptotic behavior of Lapse and Shift in the second
conformal compactification discussed in section \ref{sec:LAconf}.
In a neighborhood of $\partial M$, we may decompose $\Shift$ as 
$\Shift = \alpha \norm + \beta$. 
Then the boundary condition for $\Shift$ 
is $\alpha = 1, \beta = 0$ on $\partial M$. 

\subsection{Evolution at $\partial \aM$} 
As discussed in~\cite[\S 5]{AC}, the shear free condition is necessary in
order for the development of the data on $M$ to have a regular conformal
boundary. It is convenient to introduce the notation 
$\lambda = h^{CD}\lambda_{CD}$, $\kappa = h^{CD} K_{CD}$. We relax the
condition $\lambda \PM = 0$ used in section \ref{sec:regularity}.
 
Imposing the boundary gauge condition (\ref{eq:bdryNX}), a
calculation  shows that 
$$
\partial_t h_{AB} \PM = - ( \kappa + \lambda ) h_{AB} \PM  .
$$
In particular, $\partial_t h_{AB} \PM $ is pure trace. 
For a symmetric tensor $t_{AB}$, let 
$t_{AB}^* = t_{AB} - \half h^{CD} t_{CD} h_{AB}$ denote the  
trace free part. Define the shear tensor $\sigma_{AB}$ by 
$$
\sigma_{AB} = \lambda_{AB}^* + K_{AB}^* ,
$$
so that the shear free condition can be formulated as 
$$
\sigma_{AB} \PM = 0 .
$$
From the fact that $\partial_t h_{AB} \PM $ is 
pure trace together with the fact
that $M$ is 3--dimensional, it follows that 
$$
\partial_t \sigma_{AB} \PM = [\partial_t (K_{AB} + \lambda_{AB})]^* .
$$
Thus, if $\partial_t ( K_{AB} + \lambda_{AB}) \PM $ 
is pure trace, then the shear
free condition is conserved by the evolution. 
The evolution equations and the boundary conditions imply 
$$
\partial_t \lambda_{AB} \PM = \partial_x ( \lambda_{AB} + K_{AB} ) + \Lapse_1
K_{AB} - \half \Lie_{\Shift_1} g_{AB} \PM ,
$$
where $\Lapse_1 = \partial_x \Lapse \PM$, 
$\Shift_1 = [\partial_x , \Shift ] \PM$. 
After a lengthy calculation one finds, 
$$
\partial_t \sigma_{AB} \PM = \Lapse_1 K_{AB}^* -\half [\Lie_{\Shift_1}
g_{AB}]^* .
$$
The leading order term $\Lapse_1$ is determined by the regularity of
$\aR_{\alpha\beta}$ up to $\partial \aM$. 
Therefore the condition for preserving the shear free condition under 
evolution, 
$\partial_t \sigma_{AB}\PM = 0$ shows up as a gauge
condition for $\Shift$. Due to the shear free condition, the equation 
$$
\Lapse_1 K_{AB}^* -\half [\Lie_{\Shift_1} g_{AB}]^* \PM = 0 ,
$$ 
can be solved
by putting $\Shift_1 \PM = \Lapse_1 \norm$, 
in other words by imposing the
condition that $\Shift$ is parallel to $\norm$ 
at $\partial M$ to first order, and that $\partial_t$ is null to first order.

\section{Discussion}

In this note, I have shown how to construct solutions of the conformal
constraint equations, and 
indicated the first steps in analyzing the evolution problem for the
Einstein vacuum evolution equations, directly in the unphysical setting. It
seems worthwhile to explore gauge choices, hyperbolic reformulations,
boundary conditions etc. for the evolution equations directly in this
setting. This makes it possible to tie in to the extensive literature on
the standard form of the Einstein evolution equations. As an example, it is
interesting to look at reformulations of the Einstein equations, 
derived by adding multiples of the constraints in the evolution equations, 
which 
have improved stability of the constraints, see for example
\cite{frittelli-reula99}. It is natural to ask here if
there are such reformulations which are also better behaved at $\scri$.  

The gauge conditions that have been considered so far
in the literature on the numerical solution of the 
Einstein equations are mainly hyperbolic in
nature, and therefore suffer from well known problems such as gauge
singularities~\cite{alcubierre:masso,alcubierre:shock}.
It is known that CMC time gauge gives a well posed form of the
Einstein evolution equations. Therefore it seems natural to consider
this time gauge also for the hyperboloidal IVP.

Choosing a time gauge condition such as the CMC condition, leaves open
the choice of spatial gauge. It seems likely that it is advantageous
to work with a ``fully gauge fixed formulation'' for numerical
work. The
combination of CMC time gauge and harmonic spatial gauge gives a well
posed \index{elliptic--hyperbolic} system in case of compact Cauchy
surface~\cite{andersson:moncrief:local}. Does this gauge condition
give a well posed system also for the hyperboloidal IVP, with boundary
conditions given by a foliation of $\scri$? The defining equation for
the Shift vector field given in~\cite{andersson:moncrief:local} is
derived from the evolution equation for the metric and therefore does
not depend on the matter content of the spacetime.  This makes it
possible to apply this formulation both in the physical and unphysical
spacetime. It is an interesting problem to explore the consequences of
this gauge choice for the hyperboloidal IVP, both from the point of
view of the physical and unphysical formulation of the IVP and gauge
conditions.

\bigskip
\noindent{\bf Acknowledgements.}
This paper is based on a talk given at the Workshop on the Conformal
Structure of Space-Times held at T\"ubingen, April 2--4, 2001. I am
grateful to the organizers, J\"org Frauendiener and Helmut Friedrich,
for their hospitality and support. This work is supported in part by
the Swedish Natural Sciences Research Council (SNSRC), contract no.
R-RA 4873-307 and NSF, contract no. DMS 0104402.

\bibliography{conf}

\providecommand{\bysame}{\leavevmode\hbox to3em{\hrulefill}\thinspace}
\begin{thebibliography}{10}

\bibitem{alcubierre:shock}
Miguel Alcubierre, \emph{Appearance of coordinate shocks in hyperbolic
  formalisms of general relativity}, Phys. Rev. D (3) \textbf{55} (1997),
  no.~10, 5981--5991.

\bibitem{alcubierre:masso}
Miguel Alcubierre and Joan Mass\'o, \emph{Pathologies of hyperbolic gauges in
  general relativity and other field theories}, Phys. Rev. D (3) \textbf{57}
  (1998), no.~8, R4511--R4515.

\bibitem{ACPRL}
Lars Andersson and Piotr~T. Chru{\'s}ciel, \emph{Hyperboloidal {C}auchy data
  for vacuum {E}instein equations and obstructions to smoothness of null
  infinity}, Phys. Rev. Lett. \textbf{70} (1993), no.~19, 2829--2832.

\bibitem{AC}
\bysame, \emph{On ``hyperboloidal'' {C}auchy data for vacuum {E}instein
  equations and obstructions to smoothness of scri}, Comm. Math. Phys.
  \textbf{161} (1994), no.~3, 533--568.

\bibitem{ACdiss}
\bysame, \emph{Solutions of the constraint equations in general relativity
  satisfying ``hyperboloidal boundary conditions''}, Dissertationes Math.
  (Rozprawy Mat.) \textbf{355} (1996), 100.

\bibitem{ACF}
Lars Andersson, Piotr~T. Chru{\'s}ciel, and Helmut Friedrich, \emph{On the
  regularity of solutions to the {Y}amabe equation and the existence of smooth
  hyperboloidal initial data for {E}instein's field equations}, Comm. Math.
  Phys. \textbf{149} (1992), no.~3, 587--612.

\bibitem{andersson:iriondo:CMC}
Lars Andersson and Mirta~S. Iriondo, \emph{Existence of constant mean curvature
  hypersurfaces in asymptotically flat spacetimes}, Ann. Global Anal. Geom.
  \textbf{17} (1999), no.~6, 503--538.

\bibitem{andersson:moncrief:local}
Lars Andersson and Vincent Moncrief, \emph{Elliptic--hyperbolic systems and the
  {E}instein equations}, gr-qc/0110111, 2001.

\bibitem{frauendiener:living}
J{\"o}rg Frauendiener, \emph{Conformal infinity}, Living Rev. Relativ.
  \textbf{3} (2000), 2000--4, 92 pp. (electronic).

\bibitem{HF}
Helmut Friedrich, \emph{Cauchy problems for the conformal vacuum field
  equations in general relativity}, Comm. Math. Phys. \textbf{91} (1983),
  no.~4, 445--472.

\bibitem{frittelli-reula99}
Simonetta Frittelli and Oscar~A. Reula, \emph{Well-posed forms of the $3+1$
  conformally-decomposed {E}instein equations}, J. Math. Phys. \textbf{40}
  (1999), no.~10, 5143--5156.

\bibitem{hubner4}
Peter H{\"u}bner, \emph{How to avoid artificial boundaries in the numerical
  calculation of black hole spacetimes}, Classical Quantum Gravity \textbf{16}
  (1999), no.~7, 2145--2164.

\bibitem{hubner3}
\bysame, \emph{A scheme to numerically evolve data for the conformal {E}instein
  equation}, Classical Quantum Gravity \textbf{16} (1999), no.~9, 2823--2843.

\bibitem{hubner1}
\bysame, \emph{From now to timelike infinity on a finite grid}, Classical
  Quantum Gravity \textbf{18} (2001), no.~10, 1871--1884.

\bibitem{hubner2}
\bysame, \emph{Numerical calculation of conformally smooth hyperboloidal data},
  Classical Quantum Gravity \textbf{18} (2001), no.~8, 1421--1440.

\bibitem{penrose:rindler:II}
Roger Penrose and Wolfgang Rindler, \emph{Spinors and space-time. {V}ol. 2},
  second ed., Cambridge University Press, Cambridge, 1988, Spinor and twistor
  methods in space-time geometry.

\end{thebibliography}
\bibliographystyle{amsplain}
\end{document}